\documentclass[12pt, a4paper]{article}
\usepackage{graphicx}
\usepackage{amsmath}
\usepackage{booktabs}
\usepackage{geometry}
\usepackage{array}
\usepackage{float}
\usepackage{hyperref}
\title{A simple all-inorganic hole-only structure for trap density measurement in perovskite solar cells}

\author{Atena Mohamadnezhad$^{1}$, Alireza Fathi-Beiraghvandi$^{1}$, Mahmoud Samadpour$^{1}$\footnote{samadpour@kntu.ac.ir}}
\date{$^{1}$Department of Physics, K. N. Toosi University of Technology, Tehran, Iran}

\begin{document}

\maketitle

\begin{abstract}
One of the critical challenges in enhancing the performance of perovskite solar cells is reducing the density of trap states in the light-absorbing perovskite layer. These trap states lead to increased charge carrier recombination, thus dropping device efficiency. Space charge limited current (SCLC) analysis serves as a valuable method to study trap density, requiring structures capable of selectively transporting either electrons or holes. By analyzing current-voltage (I-V) characteristics and identifying the voltage at which the slope changes, trap density can be calculated effectively.

Traditional organic polymer hole transport layers such as Spiro-OMeTAD, PEDOT:PSS, and PTAA face challenges, including moisture instability, low charge mobility, low conductivity, and high costs. This work introduces a novel hole-only device structure utilizing inorganic materials, offering improved stability, straightforward fabrication, and reduced costs compared to conventional structures.

This device comprises a nanostructured NiO$_x$ layer, a perovskite layer, a copper indium selenide (CIS) layer, and an Au electrode on an ITO substrate. The performance of this structure is assessed by fabricating various perovskite layers under different experimental conditions. The trap density was successfully determined using the proposed hole-only device structure. Analysis of the photovoltaic properties revealed a clear correlation between trap density in the perovskite layers and the overall performance of the solar cells.
\end{abstract}

\section{INTRODUCTION}
The environmental impact and limited availability of fossil fuels have intensified efforts to develop renewable energy technologies, with solar energy conversion being a major focus. Among emerging photovoltaic technologies, perovskite solar cells (PSCs) have gained significant attention due to the exceptional properties of perovskite materials. Typically represented by the formula ABX$_3$, these materials feature a 3D framework of corner-sharing BX$_6$ octahedra (B = Pb$^{2+}$/Sn$^{2+}$; X = I$^{-}$, Br$^{-}$, or Cl$^{-}$), with larger A-site cations such as MA$^{+}$, FA$^{+}$, or Cs$^{+}$ occupying the interstitial sites. Perovskite layers are easily fabricated from solution at low temperatures and offer tunable optoelectronic properties, strong absorption across the visible--infrared spectrum, and high charge carrier mobility. However, practical implementation still faces challenges including moisture sensitivity, low thermal stability, interfacial energy mismatches, and the presence of trap states in the bulk and at surfaces.

Among these, trap states are a major factor limiting PSC performance by promoting charge carrier recombination and reducing efficiency. These defects are classified as shallow or deep based on their energy levels: shallow traps lie near the band edges, while deep traps reside within the bandgap and act as non-radiative recombination centers, causing significant carrier loss \cite{roy2020review, queisser1998defects, graetzel2017rise}. To characterize these defects, various methods have been developed, falling into three main categories: optical, electrical, and ion migration techniques \cite{srivastava2023advanced}.

Optical methods often use pump-probe setups to track carrier relaxation dynamics. Among them, time-resolved photoluminescence (TRPL) is widely used to analyze recombination via both shallow and deep traps. However, TRPL sensitivity depends on film thickness, and accurate interpretation requires complex analysis \cite{pean2020interpreting}. Electrical techniques---such as current-voltage and capacitance-voltage measurements---are widely adopted due to their ability to quantify trap densities and their energy/spatial distributions. Ion migration methods probe defect behavior under electric fields or light, particularly at interfaces and grain boundaries. Thermal admittance spectroscopy (TAS), often considered an optoelectronic technique, measures junction capacitance versus temperature to identify trap energy levels under dark or illuminated conditions. While useful for both shallow and deep traps, TAS cannot distinguish proximity to the conduction or valence bands, and slow-emission traps may remain undetected \cite{shao2014origin, lin2017matching, lin2016pi}. Among electrical techniques, deep-level transient spectroscopy (DLTS) is a sensitive method for probing a wide range of defect depths in complete devices, providing capture cross-sections for both electrons and holes. However, it may overlook minority carrier traps and struggles with shallow-level defects due to fast emission rates \cite{yang2017iodide}.

The space-charge-limited current (SCLC) method, typically applied to electron-only or hole-only devices, enables estimation of total trap densities and their activation energies from temperature-dependent measurements \cite{shi2015low, ranjan2022low, dong2015electron}. Deep-level capacitance profiling (DLCP) is another full-device technique that maps spatial and energetic distributions of traps with high resolution, though it requires complex setups and data interpretation \cite{ni2020resolving}. Among these methods, SCLC analysis stands out for its simplicity and effectiveness in characterizing trap distributions. It involves measuring current density as a function of applied voltage, offering a direct route to evaluate trap-related properties in semiconductors \cite{dahlstrom2020extraction, rohr2018exploring}.

In SCLC analysis, the current--voltage (J--V) characteristics typically exhibit four regimes: ohmic, trap-limited, trap-filling, and trap-free. In the ohmic regime, current increases linearly with voltage and is governed by intrinsic charge carriers. As voltage rises, injected carriers begin to fill trap states, marking the trap-limited regime, where current follows the Mott--Gurney law and is constrained by space charge accumulation \cite{rohr2018exploring}. Further voltage increase leads to the trap-filling regime, characterized by a sharp rise in current as deeper traps are progressively occupied. Once traps are saturated, the device enters the trap-free regime, where current again follows the Mott--Gurney law, now determined primarily by carrier mobility in a nearly trap-free environment. Although these regimes may not always be distinctly visible in experimental J--V curves, trap density can still be estimated by identifying the trap-filled limit voltage ($V_{\mathrm{TFL}}$) using established models \cite{poorkazem2018improving}.

Due to its simplicity and effectiveness, the SCLC method has been widely employed to quantify trap densities in PSCs using hole-only or electron-only structures. For instance, Gao et al. inserted a FAPbI$_x$Cl$_{3-x}$ interfacial layer between SnO$_2$ and FAPbI$_3$ via a vapor--solid reaction, reducing trap density from 5.76 $\times$ 10$^{15}$ to 4.92 $\times$ 10$^{15}$ cm$^{-3}$ and increasing PCE from 19.05\% to 22.89\% \cite{gao2025enhancing}. Ming et al. applied FAI-induced secondary grain growth to MAPbI$_3$ films, reducing trap density from 3.53 $\times$ 10$^{15}$ to 1.76 $\times$ 10$^{15}$ cm$^{-3}$, leading to improved mobility and device performance \cite{ming2024crystallization}. Wu et al. studied ion migration--induced hysteresis in PSCs and, using a FTO/PEDOT:PSS/CH$_3$NH$_3$PbI$_3$/Spiro-OMeTAD/Ag hole-only SCLC device, found trap densities ranging from 2.5 $\times$ 10$^{16}$ to 5.9 $\times$ 10$^{16}$ cm$^{-3}$, demonstrating that controlling ion migration reduces hysteresis and enhances performance \cite{wu2023effects}. Thilakan et al. compared ZnO electron transport layers (ETL) prepared by hydrothermal and sol--gel methods; the former yielded lower trap density (1.28 $\times$ 10$^{16}$ vs. 2.1 $\times$ 10$^{16}$ cm$^{-3}$) and higher PCE (18.66\% vs. 13.39\%) \cite{thilakan2023investigations}.

Qureshi et al. used a Fe$_3$O$_4$/spiro-OMeTAD hole transport bilayer, lowering trap density from 1.42 $\times$ 10$^{16}$ to 1.30 $\times$ 10$^{16}$ cm$^{-3}$ and improving charge extraction \cite{qureshi2023fe3o4}. Li et al. employed an ITO/NiO$_x$/perovskite/Spiro-OMeTAD/Au device to achieve 21\% PCE, with a trap density of 2.56 $\times$ 10$^{16}$ cm$^{-3}$ \cite{li2019niox}. Zhu et al. achieved 23.25\% efficiency by incorporating DMIMPF$_6$ ionic liquid to reduce surface traps to 6.04 $\times$ 10$^{15}$ cm$^{-3}$ \cite{zhu2021high}. Yuan et al. used CsPbI$_2$Br perovskite and SCLC analysis with an FTO/TiO$_2$/perovskite/PCBM/Ag structure to investigate trap states \cite{yuan2018surface}. Gao et al. incorporated 5 wt\% benzophenone to improve monocrystalline perovskite film morphology, reducing trap density to 8.9 $\times$ 10$^{15}$ cm$^{-3}$ with an ITO/PCBM/perovskite/PCBM/Ag configuration \cite{gao2018spin}. Wu et al. doped perovskite films with Eu$^{2+}$ to enhance stability and reduce trap density to 2.67 $\times$ 10$^{15}$ cm$^{-3}$, verified by SCLC in an FTO/SnO$_2$/perovskite/PCBM/Ag structure \cite{wu2018ch3nh3pb1}. Han et al. studied cesium doping in lead(II)-acetate-based PSCs, observing an efficiency increase from 14.1\% to 15.57\% and a trap density of 3.6 $\times$ 10$^{16}$ cm$^{-3}$ measured via SCLC in an ITO/PTAA/perovskite/PTAA/Ag configuration \cite{han2021cesium}.

As evident from the aforementioned studies, as well as other valuable works \cite{xu2019high, wang2019interface, cao2019flexible, zhang2018solution, yao2020dimensionality, jiang2020in, liu2020promoting, kim2020high, yang2020efficient, liu2019fabrication, li2020efficient}, organic charge transport materials are widely employed in perovskite solar cells and in electron-/hole-only devices. While these materials have enabled high power conversion efficiencies, their inherent limitations have prompted researchers to explore inorganic alternatives as potential charge transport layers. For instance, Spiro-OMeTAD---one of the most commonly used hole transport materials---exhibits relatively low hole mobility and presents challenges in forming uniform thin films. Moreover, it suffers from poor stability under humid conditions \cite{hua2018composite}. Another widely used HTL, PEDOT:PSS, is hydrophilic, making it highly sensitive to moisture. Additionally, its acidic nature can lead to undesirable chemical interactions with the perovskite layer over time, contributing to device degradation \cite{xu2017functional}. PTAA is another frequently employed hole transport material in photovoltaic devices; however, its low intrinsic mobility necessitates doping to achieve acceptable performance. Unfortunately, such doping can adversely affect the device by increasing hysteresis \cite{xu2021conjugated}. On the electron transport side, a variety of ETLs have been explored. SnO$_2$, although commonly used, has drawbacks such as low mobility, pronounced hysteresis, and a high density of trap states when used in its pristine form \cite{luan2019high}.

In contrast, inorganic materials such as NiO$_x$ offer several advantages. NiO$_x$ exhibits high hole mobility \cite{yin2019nickel}, minimal absorption in the visible region---thus not interfering with light harvesting---and excellent environmental and moisture stability \cite{deshpande2016structural}. Furthermore, its valence band aligns well with that of perovskite, facilitating efficient hole extraction and making it a promising candidate for improving the long-term stability and performance of perovskite solar cells. In this study, we introduce a new device structure based entirely on inorganic layers for SCLC analysis, utilizing the ITO/NiO$_x$/Perovskite/CIS/Au configuration. In this hole-only device, indium tin oxide (ITO) serves as the substrate, while NiO$_x$ and copper indium selenide (CIS) function as inorganic hole transport layers (HTLs), sandwiching the perovskite absorber. A gold (Au) electrode is deposited on top of the CIS layer. CIS, an emerging inorganic HTL, has recently been applied in perovskite solar cells with promising results \cite{khorasani2019optimization}. It offers high conductivity, excellent stability, and a valence band energy level well-aligned with that of perovskite, making it a suitable material for efficient hole extraction.

By employing inorganic NiO$_x$ and CIS as HTLs, we successfully fabricated stable and cost-effective hole-only devices for SCLC measurements. To validate the effectiveness of this new device configuration, several PSCs were fabricated, and their performance was correlated with the trap densities obtained from the proposed structure. The results demonstrate that this all-inorganic device provides a simple, stable, and reliable platform for quantifying trap densities in perovskite layers, thereby contributing to the development of more efficient and durable perovskite solar cells.

\section{Experimental Section}
\subsection{Materials and synthesizing methods}
Fluorine-doped tin oxide (FTO) coated glass substrates were first patterned using zinc powder and diluted hydrochloric acid. The patterned substrates were then sequentially cleaned by ultrasonic treatment in Hellmanex solution, deionized water, ethanol, acetone, and isopropanol, each for 10 minutes. Following the cleaning process, the substrates were annealed at 450\,$^\circ$C for 15 minutes and subsequently treated with UV-ozone to eliminate any remaining surface contaminants.

To fabricate the compact TiO$_2$ blocking layer, a 0.15 M solution of titanium isopropoxide (TTIP) in ethanol was spin-coated onto the FTO substrates at 2000 rpm for 30 seconds, followed by annealing at 500\,$^\circ$C for 1 hour. Next, a commercial TiO$_2$ paste was diluted in ethanol at a weight ratio of 1:5.5 and spin-coated at 4000 rpm for 30 seconds over the compact TiO$_2$ layer. The resulting mesoporous TiO$_2$ film was sintered at 500\,$^\circ$C for 30 minutes. Prior to perovskite deposition, the FTO/compact TiO$_2$/mesoporous TiO$_2$ substrates were exposed to UV-ozone treatment for 5 minutes to ensure surface cleanliness and optimal wettability.

To prepare the perovskite precursor solution, lead iodide (PbI$_2$, 1.1 M), formamidinium iodide (FAI, 1.0 M), methylammonium bromide (MABr, 0.2 M), cesium iodide (CsI, 0.05 M), and lead bromide (PbBr$_2$, 0.22 M) were dissolved in a mixed solvent of anhydrous dimethylformamide (DMF) and dimethyl sulfoxide (DMSO) at a volume ratio of 4:1. The resulting solution was deposited onto the prepared mesoporous TiO$_2$ layer via a two-step spin-coating process. The first step was carried out at 1000 rpm for 10 seconds, followed by a second spin at 4000 rpm for 20 seconds. During the final 5 seconds of the second step, 250 $\mu$L of chlorobenzene was dropped onto the spinning substrate as an anti-solvent to induce rapid crystallization.

Modified perovskite layers were also fabricated using an alternative anti-solvent, prepared by mixing tetraethyl orthosilicate (TEOS) with chlorobenzene at a volume ratio of 3:400. These perovskite films were subsequently annealed at 100\,$^\circ$C for 45 minutes to enhance crystallinity. A thin layer of Spiro-OMeTAD was then deposited as the HTL by spin coating at 500 rpm for 30 seconds. The HTL precursor solution was composed of 72.3 mg of Spiro-OMeTAD dissolved in 1 mL of chlorobenzene, with the addition of 28.8 $\mu$L of 4-tert-butylpyridine and 17.5 $\mu$L of a 1.8 M LiTFSI solution in acetonitrile. Finally, gold electrodes ($\sim$100 nm thick) were thermally evaporated to complete the device fabrication. For the hole-only devices, an ITO/NiO$_x$/Perovskite/CIS/Au architecture was employed, where ITO functioned as both the substrate and hole-collecting electrode. NiO$_x$ and copper indium selenide (CIS) served as inorganic hole transport layers, prepared according to methods previously reported in the literature \cite{khorasani2019optimization, saki2019effect}.

\subsection{Characterization Methods}
X-ray diffraction (XRD) patterns were obtained with a Bruker D8 advance diffractometer equipped with Cu K$\alpha$ radiation ($\lambda$ = 1.54 \AA). Surface morphology was analyzed via scanning electron microscopy (SEM) using a MIRA3 TESCAN field emission microscope. The optical properties of the perovskite films were evaluated using a UVS-2500 spectrophotometer (PHYSTEC). Photocurrent--voltage (J--V) measurements were carried out under standard AM 1.5G solar illumination (1000 W/m$^2$) using a Sharif Solar SIM-1000 solar simulator. A Keithley 2400 digital source meter was used to record the J--V characteristics, and the devices were masked during measurement to define an active area of 0.16 cm$^2$. Space-charge-limited current (SCLC) measurements were conducted on hole-only devices with the configuration ITO/NiO$_x$/Perovskite/CIS/Au. For steady-state photoluminescence (PL) analysis, samples were excited with a 405 nm laser source, and the emission spectra were collected using an Avaspec 2048 TEC spectrophotometer.

\section{RESULTS AND DISCUSSION}
Figure 1 presents the SEM image of the perovskite layer. As observed, the film predominantly exhibits large grain sizes, although a few smaller grains are also present.

\begin{figure}[H]
\centering
\includegraphics[width=0.5\textwidth]{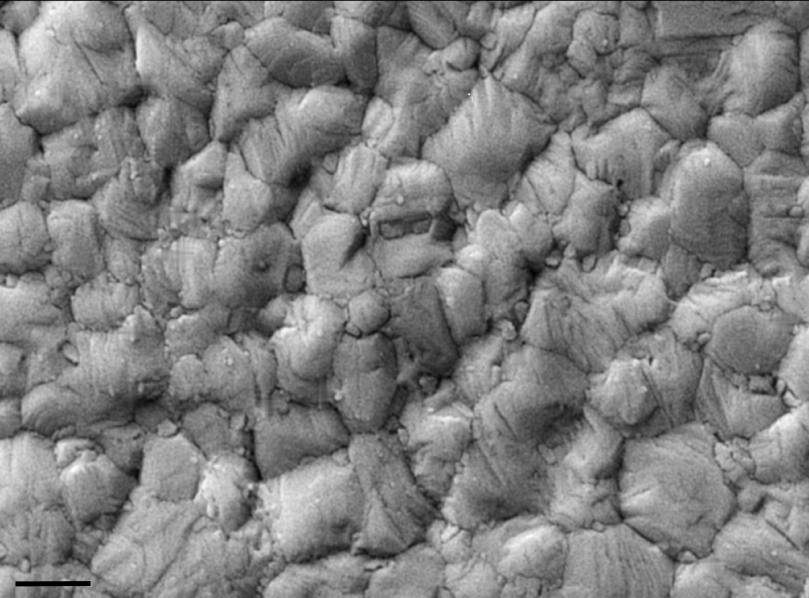}
\caption{SEM image of perovskite (PS) layer. The scale bar represents 500 nm.}
\end{figure}

The crystalline structures of the pristine and TEOS-modified perovskite films were analyzed using X-ray diffraction (XRD), as shown in Figure 2. The diffraction peaks at 14.18$^\circ$, 20.07$^\circ$, 24.60$^\circ$, 28.46$^\circ$, 31.88$^\circ$, 35.02$^\circ$, 40.62$^\circ$, and 43.21$^\circ$ correspond to the characteristic planes of the Cs$_{0.05}$(MA$_{0.17}$FA$_{0.83}$)$_{0.95}$Pb(I$_{0.83}$Br$_{0.17}$)$_3$ perovskite structure \cite{saliba2016cesium}. The full width at half maximum (FWHM) of the peak at 40.62$^\circ$ was 0.11 for both samples, indicating comparable crystallinity. The crystallite size, estimated using the Scherrer equation from the most intense peak, was approximately 77 nm. The consistent peak positions in both films suggest that the crystal structure is preserved upon TEOS modification, likely due to similar anti-solvent-induced crystallization pathways. In addition to the perovskite and FTO peaks, a diffraction peak at 12.7$^\circ$ corresponding to residual PbI$_2$ was observed in both samples, likely resulting from incomplete precursor conversion under ambient deposition conditions. While trace amounts of PbI$_2$ can passivate defects and improve performance, excessive PbI$_2$ may be detrimental due to its suboptimal optoelectronic properties \cite{chen2019mechanism}. Following the addition of TEOS to the anti-solvent, a silica-containing layer is expected to form on the perovskite surface via hydrolysis; however, no diffraction peaks associated with crystalline silica were detected, suggesting the silica exists in an amorphous form.

\begin{figure}[H]
\centering
\includegraphics[width=0.5\textwidth]{2.png}
\caption{X-ray diffraction (XRD) patterns of pristine and TEOS-modified perovskite films deposited on FTO-coated glass substrates. Diffraction peaks corresponding to the perovskite phase, residual PbI$_2$, and FTO are labeled with $\alpha$, *, and \#, respectively.}
\end{figure}

To investigate the optical properties, UV--vis absorption spectra of the pristine and TEOS-modified perovskite films were measured, as shown in Figure 3. The spectra exhibit only minor differences, indicating that the TEOS modification does not significantly alter the optical absorption characteristics. Tauc plot analysis revealed optical bandgaps of approximately 1.58 eV for both films. These findings suggest that the TEOS treatment has a negligible effect on the optical properties of the perovskite layer. To assess the impact of this modification on device performance, perovskite solar cells (PSCs) were fabricated using both pristine and TEOS-modified films.

\begin{figure}[H]
\centering
\includegraphics[width=0.5\textwidth]{3.png}
\caption{UV--vis absorption spectra of pristine and TEOS-modified perovskite layer.}
\end{figure}

Figure 4 shows the J--V curves of the pristine and TEOS-modified perovskite solar cells, with the key photovoltaic parameters summarized in Table 1. The pristine devices exhibited a $V_{\mathrm{oc}}$ of 1.01 V, $J_{\mathrm{sc}}$ of 22.46 mA$\cdot$cm$^{-2}$, fill factor (FF) of 62.60\%, and a power conversion efficiency (PCE) of 14.06\%. These relatively modest values are attributed to fabrication under ambient conditions, which generally yield lower efficiencies compared to triple-cation perovskite cells processed in inert environments such as N$_2$ or Ar \cite{saliba2016cesium}. In comparison, the TEOS-modified devices demonstrated enhanced performance, with a higher $J_{\mathrm{sc}}$ of 23.39 mA$\cdot$cm$^{-2}$, $V_{\mathrm{oc}}$ of 1.03 V, and a substantially improved FF of 71.26\%, resulting in an increased PCE of 17.16\%. The comparison between the J--V curves and Table 1 clearly illustrates the superior performance of the modified cells, primarily due to the significant improvement in fill factor, while $V_{\mathrm{oc}}$ and $J_{\mathrm{sc}}$ exhibit modest enhancements.

\begin{figure}[H]
\centering
\includegraphics[width=0.6\textwidth]{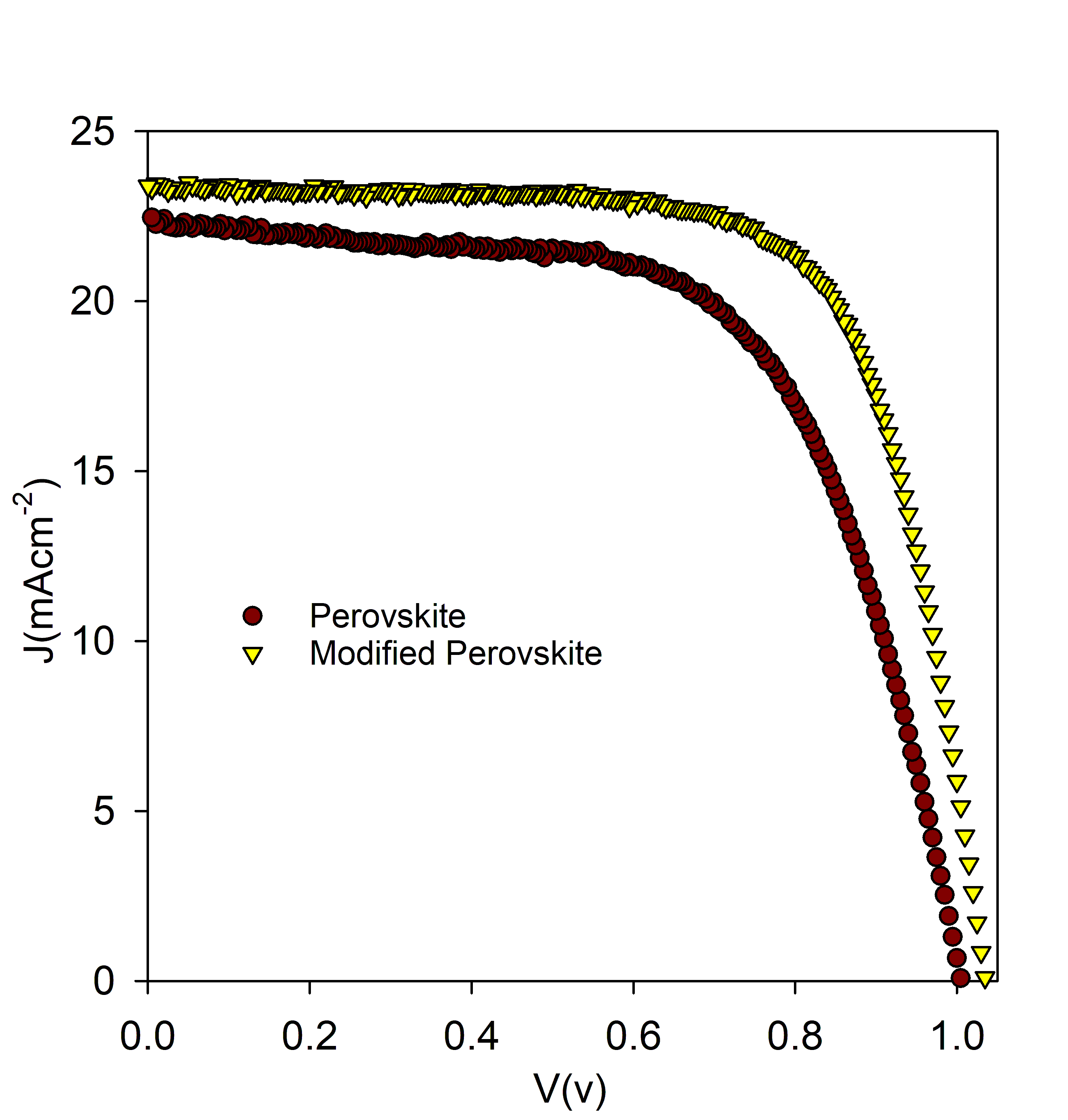}
\caption{J--V characteristics of perovskite solar cells incorporating pristine and TEOS-modified perovskite layers.}
\end{figure}

\begin{table}[H]
\centering
\caption{Photovoltaic parameters of the solar cells, including short-circuit current density ($J_{\mathrm{sc}}$), open-circuit voltage ($V_{\mathrm{oc}}$), fill factor (FF), and power conversion efficiency (PCE) for pristine (normal) and TEOS-modified devices.}
\begin{tabular}{cccccc}
\toprule
Cell type & $V_{\mathrm{oc}}$ (V) & $J_{\mathrm{sc}}$ (mA/cm$^2$) & FF (\%) & PCE (\%) \\
\midrule
Pristine Cell & 1.00 & 22.46 & 62.60 & 14.06 $\pm$ 0.22 \\
Modified Cell & 1.03 & 23.39 & 71.26 & 17.16 $\pm$ 0.13 \\
\bottomrule
\end{tabular}
\end{table}

Steady-state photoluminescence (PL) measurements were conducted on both pristine and TEOS-modified perovskite (PS) layers to investigate the origin of the enhanced photovoltaic performance observed in the modified devices. Figure 5 presents the PL spectra of the pristine and TEOS-modified Cs$_{0.05}$(MA$_{0.17}$FA$_{0.83}$)$_{0.95}$Pb(I$_{0.83}$Br$_{0.17}$)$_3$ films. Both samples exhibit a PL peak centered at 778 nm; however, the TEOS-modified layer shows a markedly higher PL intensity. This enhancement suggests a significant reduction in trap state density, which typically contributes to non-radiative recombination. The unchanged peak position, along with the structural and optical consistency observed in Figures 1 to 3, indicates that the bulk properties of the perovskite layers remain unaffected by the TEOS treatment.

\begin{figure}[H]
\centering
\includegraphics[width=0.5\textwidth]{5.png}
\caption{Photoluminescence emission spectra of pristine Cs$_{0.05}$(MA$_{0.17}$FA$_{0.83}$)$_{0.95}$Pb(I$_{0.83}$Br$_{0.17}$)$_3$ perovskite (PS) films and TEOS-modified PS films.}
\end{figure}

The trap state density was evaluated using space-charge-limited current (SCLC) measurements, as shown in Figure 6. These measurements were carried out on hole-only devices with the architecture ITO/NiO$_x$/Perovskite/CIS/Au, illustrated in the inset of Figure 6.

\begin{figure}[H]
\centering
\includegraphics[width=0.5\textwidth]{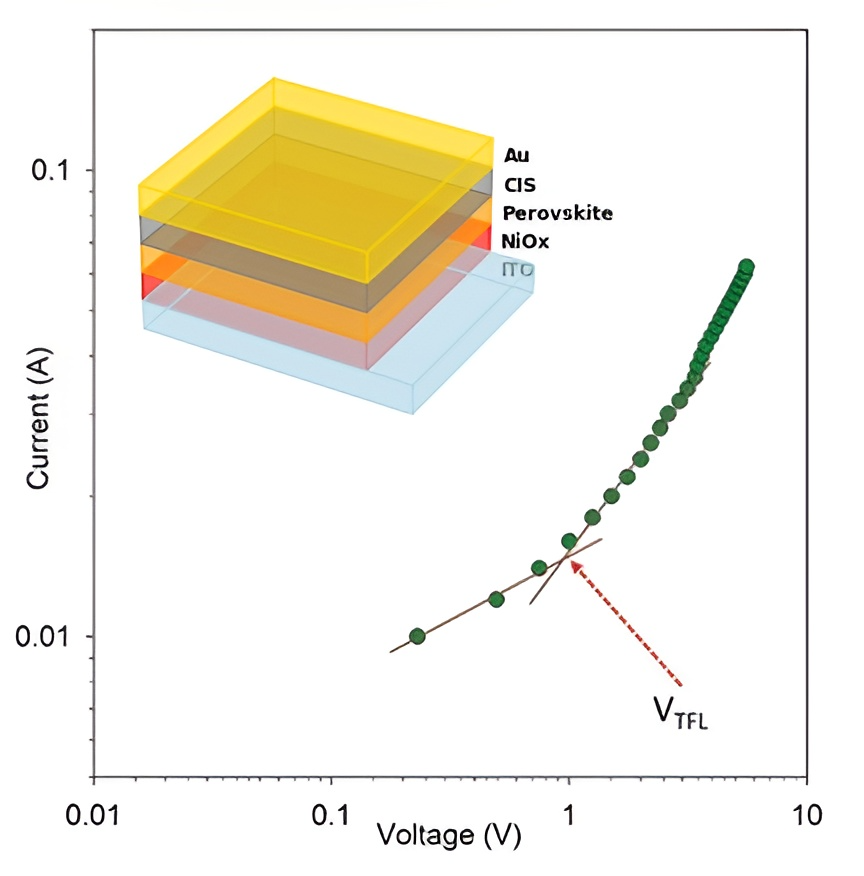}
\caption{Typical current-voltage characteristics of the hole-only devices with ITO/NiO$_x$/Perovskite/CIS/Au structure shown in the inset.}
\end{figure}

The trap density ($D_{\mathrm{trap}}$) was determined from the trap-filled limit voltage ($V_{\mathrm{TFL}}$) using the equation: $V_{\mathrm{TFL}} = 0.5 e D_{\mathrm{trap}} L^2 \varepsilon^{-1} \varepsilon_0^{-1}$, where $e$ is the elementary charge, $\varepsilon_0$ is the vacuum permittivity, $\varepsilon$ is the relative dielectric constant of the perovskite (28.8), and $L$ is the thickness of the perovskite layer \cite{poorkazem2018improving, bube1962trap, kumar2020large}. The extracted trap densities for both pristine and TEOS-modified devices are summarized in Table 2 where each measurement was conducted on at least three devices.

\begin{table}[H]
\centering
\caption{Trap densities of pristine and modified devices (each measurement was conducted on at least three pristine/modified devices).}
\begin{tabular}{ccc}
\toprule
Cell type & Sample No & $N_t$ (cm$^{-3}$) \\
\midrule
Pristine & 1 & 1.63 $\times$ 10$^{16}$ \\
Pristine & 2 & 1.69 $\times$ 10$^{16}$ \\
Pristine & 3 & 1.74 $\times$ 10$^{16}$ \\
Modified & 1 & 1.41 $\times$ 10$^{16}$ \\
Modified & 2 & 1.46 $\times$ 10$^{16}$ \\
Modified & 3 & 1.49 $\times$ 10$^{16}$ \\
\bottomrule
\end{tabular}
\end{table}

The SCLC results clearly demonstrate that TEOS modification of the perovskite layer leads to a reduced trap density compared to pristine samples. This reduction is attributed to the presence of amorphous silica, formed through TEOS hydrolysis, which likely passivates surface trap states within the perovskite film \cite{guan2019employing}. The lower trap density contributes to the improved fill factor and, consequently, the higher power conversion efficiency observed in the modified devices, as shown in Table 1. Moreover, the consistency of results across multiple SCLC measurements highlights the reliability and accuracy of the all-inorganic hole-only device structure employed for trap density evaluation.

\section{Conclusions}
In this study, a simple all-inorganic hole-only device was developed to assess trap state density in perovskite layers. To validate the effectiveness of this structure, solar cells were fabricated using both pristine and silica-modified perovskite films. The results revealed that while the crystalline structure and optical properties remained largely unaffected by the modification, the trap state density was significantly reduced. This reduction, likely due to the passivating effect of amorphous silica, led to a marked improvement in fill factor by suppressing charge recombination. Overall, the proposed hole-only device provides a reliable and straightforward platform for monitoring defect densities, offering valuable insights for improving perovskite layer quality and advancing the performance of perovskite solar cells.

\bibliographystyle{unsrt}
\bibliography{references}

\end{document}